\documentclass[usenatbib]{mnras}

\usepackage{mathptmx}

\usepackage[T1]{fontenc}
\usepackage{ae,aecompl}

\usepackage{graphicx}	

\title[Redshift, metallicity and size of two extended dIrrs]{Redshift, metallicity and size of two extended dwarf Irregular galaxies. A link between dwarf Irregulars and ultra diffuse galaxies?\thanks{Based on data acquired using the Large Binocular Telescope (LBT).
The LBT is an international collaboration among institutions in the United
States, Italy, and Germany. LBT Corporation partners are The University of
Arizona on behalf of the Arizona university system; Istituto Nazionale di
Astrofisica, Italy; LBT Beteiligungsgesellschaft, Germany, representing the
Max-Planck Society, the Astrophysical Institute Potsdam, and Heidelberg
University; The Ohio State University; and The Research Corporation, on behalf
of The University of Notre Dame, University of Minnesota and University of
Virginia.}}

\author[M. Bellazzini et al.]{
M. Bellazzini$^{1}$\thanks{E-mail: michele.bellazzini@oabo.inaf.it},
V. Belokurov$^{2}$,
L. Magrini$^{3}$,
F. Fraternali$^{4,5}$,
V. Testa$^{6}$, 
G. Beccari$^{7}$,\newauthor
A. Marchetti$^{8,9}$,
R. Carini$^{6}$
\\
$^1$ INAF - Osservatorio Astronomico di Bologna, Via Ranzani 1, 40127 Bologna, Italy\\
$^2$Institute of Astronomy, University of Cambridge, Madingley Rd, Cambridge CB3 0HA, UK\\
$^3$INAF - Osservatorio Astrofisico di Arcetri, Largo E. Fermi 5, 50125 Firenze, Italy\\
$^4$Dipartimento di Fisica \& Astronomia, Universit\`a degli Studi di Bologna, Viale Berti Pichat, 6/2, I - 40127 Bologna, Italy\\
$^5$Leiden Observatory, Leiden University, Postbus 9513, 2300 RA, Leiden, The Netherlands\\
$^6$INAF - Osservatorio Astronomico di Roma, via Frascati 33, 00040 Monteporzio, Italy\\ 
$^7$European Southern Observatory, Karl-Schwarzschild-Strasse 2, 85748 Garching bei M\"unchen, Germany\\ 
$^8$Universit\`a degli Studi di Milano, via G. Celoria 16, 20130 Milano, Italy  \\          
$^9$ INAF - Osservatorio Astronomico di Brera, via Brera 28, 20122 Milano, via E. Bianchi 46,
             23807 Merate, Italy\\}

\date{Accepted XXX. Received YYY; in original form ZZZ}

\pubyear{2016}

\begin{document}
\label{firstpage}
\pagerange{\pageref{firstpage}--\pageref{lastpage}}
\maketitle

\begin{abstract}
We present the results of the spectroscopic and photometric follow-up of two field galaxies that were selected as possible stellar counterparts of local high velocity clouds. Our analysis shows that the 
two systems are distant (D$>20$~Mpc) dwarf irregular galaxies unrelated to the local 
H~I clouds. However, the newly derived distance and structural parameters reveal 
that the two galaxies have luminosities and effective radii very similar to the recently identified ultra diffuse galaxies (UDGs). At odds with classical UDGs, they are remarkably isolated, having no known giant galaxy within $\sim 2.0$~Mpc.
Moreover, one of them has a very high gas content compared to galaxies of similar stellar mass, with a H~I 
to stellar mass ratio 
$M_{\rm HI}/M_{\rm \star}\sim 90$, typical of {\em almost-dark} dwarfs.
Expanding on this finding, we show that extended dwarf irregulars overlap the distribution of UDGs in the $M_V$ vs. log~$r_{\rm e}$ plane and that the sequence including dwarf spheroidals, dwarf irregulars and UDGs appears as continuously populated in this plane.  
This may suggest an evolutionary link between dwarf irregulars and UDGs.
\end{abstract}

\begin{keywords}
ISM: H{\sc ii} regions --- galaxies: dwarf --- galaxies: star formation
\end{keywords}



\section{Introduction}
The complete census of dwarf galaxies in the Local Group (LG, and in the Local Volume) is a key observational enterprise in these decades, closely tied to the solution of the long-standing {\em missing satellites problem} \citep[][and references therein]{moore,sawa16}. 
The recent discovery of the nearby (D=1.7~Mpc) faint ($M_V=-9.4$) star-forming dwarf galaxy Leo~P \citep{leop_1,leop_lbt} has opened a new road for the identification of local dwarfs.
Leo~P was found as the stellar counterpart of a very compact high velocity cloud (CHVC) of neutral Hydrogen identified in the ALFALFA HI survey \citep{giova07}, thus suggesting that some of the missing dwarfs in the LG and its surroundings could be hidden within similar CHVCs. 
These dwarfs may be the gas-rich star-forming counterparts of the quiescent ultra faint dwarfs (UFD) that have been found in relatively large numbers as stellar overdensities in panoramic imaging surveys \citep[see, e.g.,][and references therein]{belo,kopo15}. 
Indeed there are models within the $\Lambda$-cold dark matter (CDM) scenario predicting that a large number of small DM haloes \citep[$M\la 10^9~M_{\rm \sun}$, mini-haloes, after][]{ricotti} should have had their star formation inhibited or quenched by global or local feedback effects (e.g., re-ionization, supernova feedback, ram-pressure stripping), thus leading to a population of gas-rich dwarfs with low or null stellar content \citep{ricotti,korfree}. 

The only possibility to confirm these systems as real galaxies and to gauge their distances is to find a stellar population associated with the HI clouds and indeed several teams followed up the CHVCs proposed by the ALFALFA \citep{adams} and GALFA-HI \citep{galfa} surveys as candidate local (D$\le 3.0$~Mpc) mini-haloes.
\citet[][B15a hereafter]{pap1}, within the SECCO survey\footnote{http://www.bo.astro.it/secco/}, obtained deep and homogeneous imaging of 25 of the ALFALFA candidates from A13, finding only one confirmed stellar counterpart, the very faint star-forming system SECCO~1, likely located in the Virgo cluster \citep{secco1}. \citet{sand15}, searching several public image archives, were able to confirm SECCO~1 and discovered four additional counterparts in the GALFA-HI sample, all of them with D$\ga 3.0$~Mpc \citep[see also][]{tolle}. 
\citet[][J15 hereafter]{bdd} adopted a different approach, searching for small groupings of blue stars within the SDSS catalogue and identifying $\sim 100$ interesting candidates. 
The follow-up of 12 of them revealed a population of faint, blue, metal-poor low surface brightness (LSB) dwarfs in the distance range 5~Mpc$\la {\rm D}\la$~120~Mpc, six of them associated with HI clouds listed in \citet{giova07}. J15 defined the newly found systems as blue diffuse dwarf (BDD) galaxies.
Apparently we are beginning to scratch the surface of a population of LSB star-forming dwarfs that went undetected until now, although they are not found in the Local Volume \citep{sand15,pap2}. 

Within this context, we have selected mini-halo candidates from the GASS HI survey \citep{McClure}. 
Unlike previous searches, which only looked at velocities very different from Galactic emission  (HVCs with $|v_{\rm \rm dev}|>90$ km~s$^{-1}$, $v_{\rm \rm dev}$ being the $deviation$ velocity with respect to a regularly rotating Galactic disc), we have explored the range of lower velocities, typical of intermediate velocity clouds (IVCs, $30<|v_{\rm \rm dev}|<90$ km~s$^{-1}$).
We detected HI sources using the code $^{\rm 3D}$BAROLO \citep{barolo} and applying selection criteria on their size and velocity width to minimize the contamination from Galactic clouds.
This left us with a sample of about one thousand best candidates, presumably with a very high degree of contamination by Galactic sources, which we searched for stellar counterparts in SDSS \citep{dr9}, ATLAS \citep{atlas} and DES \citep{des} images.
The process of visual inspection of available images around the positions of the clouds led to the selection of two promising candidates. These are blue LSB galaxies whose apparent diameters are fully compatible with being located within 
$\sim 3.0$~Mpc from us; moreover they are not completely unresolved, displaying a few blue compact sources resembling HII regions. 
Unfortunately, the spatial resolution of GASS is about 16 arcmin and it does not allow an association with certainty, hence spectroscopic follow-up is required.

Here we present the results of this follow-up, ultimately resulting in the rejection of the association of both the candidate stellar counterparts with the local gas clouds, since they are located at distances larger than 40~Mpc. Still, our observations provide the first redshift and metallicity estimates for these galaxies, which are useful for future studies, and reveal their remarkably large size, given their total luminosity. The latter feature lead us to note that the most luminous dwarf Irregular galaxies (dIrr) display structural parameters (sizes, integrated magnitudes, S\'ersic indices, and surface brightnesses) overlapping the range inhabited by the newly discovered ultra diffuse galaxies \citep[UDGs;][]{udg,koda}, suggesting a possible relation between the two classes of stellar systems.

While these observations are not part of the SECCO survey, they are strongly related and for this reason we adopt the SECCO nomenclature to name the two dwarfs considered here. In particular, following \citet{pap2} we call them SECCO-dI-1 and SECCO-dI-2 (where dI = dwarf Irregular), abbreviated as SdI-1
and SdI-2.

   \begin{figure*}
   \centering
   \includegraphics[width=0.8\textwidth]{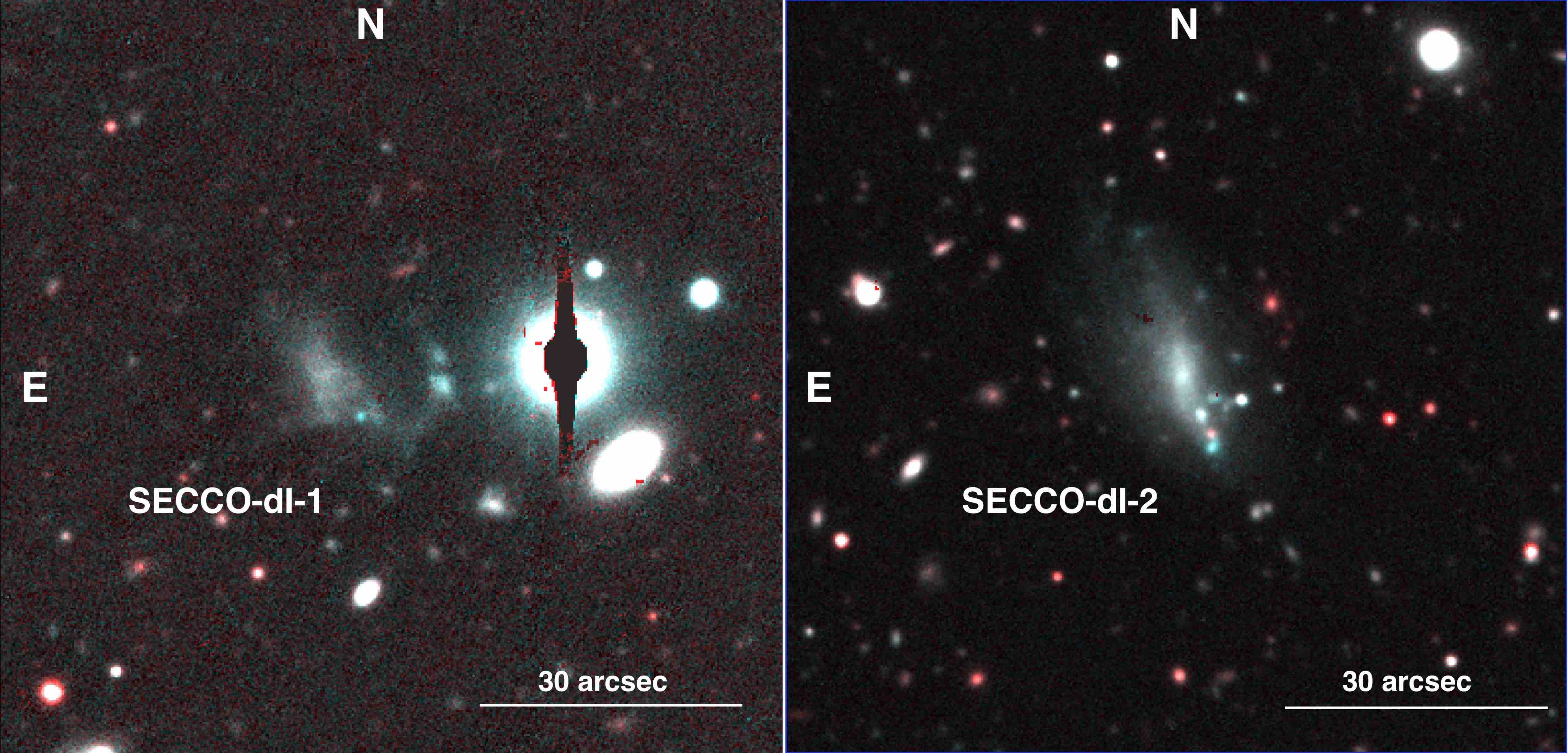}
     \caption{RGB color images of SECCO-dI-1 (left panel) and SECCO-dI-2 (right panel) obtained from our stacked LBC-RED images, using the i band for R and the r band for G and B. The scale and the orientation are shown.}
        \label{imaC1C2}
    \end{figure*}


\section{Observations and data reduction}

All the observations have been obtained under clear sky, during the night of March 3, 2016 with the Large Binocular Telescope (LBT) on Mt~Graham (AZ),  used in 
pseudo-binocular mode, i.e. with different instruments operating simultaneously in the two channels of the telescope.

The main programme was performed using the low resolution spectrograph MODS-1 \citep{mods}, aligning a $1.2\arcsec$-wide long-slit  along the major axes of the two targets. With this set-up and a dichroic, MODS-1 provides a blue spectrum covering $3200\AA<\lambda<5650\AA$, and a red spectrum covering $5650\AA<\lambda<10000\AA$ at a spectral resolution $\frac{\lambda}{\Delta \lambda}\sim 1100$. Three $t_{\rm exp}=1200$~s exposures per target were acquired. 
SDSS images were used to define the slit position and orientation.
The spectra were corrected for bias and flat-field, sky-subtracted, wavelength calibrated, then extracted and combined into flux-calibrated summed spectra using the pipeline developed at the Italian {\it LBT Spectroscopic Reduction Center}\footnote{http://lbt-spectro.iasf-milano.inaf.it}.

Simultaneously, we got deep r$_{\rm \rm SDSS}$ and i$_{\rm \rm SDSS}$ (hereafter r and i, for brevity) band imaging with LBC-R, the red channel of a pair of twin wide field ($\simeq 23\arcmin\times23\arcmin$) cameras \citep{lbc}. For each target, we got $9\times 200$~s exposures per filter, dithered to minimise the effect of bad pixels and to cover the inter-chip gaps of LBC-R. 
The reduction of the LBC images was performed with the specific pipeline developed at INAF-OAR (Paris et al. in preparation). The individual raw images were first corrected for bias and flat-field, and then background-subtracted. After astrometric calibration, they were combined into single r- and i-band stacked images with the SWarp software \citep{swarp}. In the following we will analyse these stacked and sky-subtracted images. The 5-$\sigma$ level over the background measured on the images corresponds to an i-band surface brightness of $\simeq$27.8~mag/arcsec$^2$, in line with the limits typically obtained with LBC images reduced in the same way \citep[see][]{pap2}.
The photometric calibration was obtained with hundreds of stars 
in common with the Sloan Digital Sky Survey - Data Release 9 \citep[SDSS-DR9,][]{dr9}.
In Fig.~\ref{imaC1C2} we show postage-stamp images of the two target galaxies from our LBC-R images.

\subsection{The targets}
\label{targ}

The main properties of the target galaxies are summarised in Table~\ref{mean}. SdI-1 has a determination of redshift $z=0.025988$ from H~I by \citet{rob04} that places it far beyond the realm of local mini-haloes \citep[D$\le 3$~Mpc;][]{adams}. However a negative redshift ($z=-0.05545$) was also reported from optical spectroscopy \citep{liske03}\footnote{In the Millennium Galaxy Catalog \tt http://www.hs.uni-hamburg.de/jliske/mgc/}. While extremely implausible, this may suggest an improper association between the H~I detection by \citet{rob04} and the stellar counterpart. Given the identification of an ICV along the same line of sight, a new attempt to get a reliable velocity from an optical spectrum was worth doing. \citet{rob04} also reported an integrated magnitude of B=19.2 and an effective radius $r_{\rm e}=4\arcsec$ (no uncertainty reported) from single-band photometry much shallower than the one we obtained with LBC.

SdI-2 lacks any redshift, size and optical magnitude estimate. There are two {\it GALEX} sources projected onto the main body of the galaxy, GALEXMSC J114433.79-005200.0 classified as a UV source, and GALEXASC J114433.60-005203.0 classified as a galaxy.  

Our deep images (see Fig.~\ref{imaC1C2}) reveal two remarkably elongated irregular blue galaxies, with some compact knots superimposed. SdI-1 may be interacting with some smaller companions, located $\sim 15\arcsec$ westward of its center. The main body of SdI-2 is surrounded by a very low SB asymmetric halo, more extended in the North-East direction. A fluffy nucleus is visible at the very center of SdI-2, which can be perceived also in the light profile (as a slight change of slope in the innermost 2$\arcsec$, see below). While SdI-2 displays and an overall elliptical shape, the apparent morphology of SdI-1 is remarkably irregular independently of the adopted image cuts.

\begin{table*}
  \begin{center}
  \caption{Mean Properties of the targets}
  \label{mean}
  \begin{tabular}{lccc}
\hline
                &  SECCO-dI-1 & SECCO-dI-2 & note \\
\hline
other name      & B115324.62+001916.3     &           & \citet{apmuk}  \\
other name      & MGS sure 21     &           & \citet{rob04}  \\
other name      & MGC95504                &           & \citet{liske03} \\
other name      &                 & GAMA 535033          & \citet{liske15} \\
RA$_{\rm  J2000}$              & 11:55:58.5 & 11:44:33.8  &  from LBC images     \\
Dec$_{\rm  J2000}$             &+00:02:36.3 &-00:52:00.9  &  from LBC images     \\
V$_r$ [km~s$^{-1}$]    & $7829\pm 50$ & $2549\pm 50$          & heliocentric      \\
V$_r^{\rm HI}$ [km~s$^{-1}$]    & $7791\pm 13$ &           & heliocentric from \citet{rob04}      \\
V$_{\rm 3K}$ [km~s$^{-1}$] & $8152\pm 25^a$ & $2914\pm 60$       &  3K (NED)    \\
D [Mpc]         & $112\pm 8$ & $40\pm 1$          &  $H_0=73$~km~s$^{-1}$~Mpc$^{-1}$     \\
$r_{\rm int}$    &  $19.7\pm 0.1$    & $18.8\pm 0.1$          &  total integrated magnitude$^b$     \\
$i_{\rm int}$    &  $19.8\pm 0.1$    & $18.7\pm 0.1$          &  total integrated magnitude$^b$     \\
$M_V$    &  $-15.7\pm 0.2$    & $-14.1\pm 0.2$          &  absolute integrated V magnitude$^c$     \\
$\langle \mu_V\rangle_{\rm e}$    &  $24.2\pm 0.1$    & $24.3\pm 0.1$          & [mag/arcsec$^2$]$^b$    \\
$\mu_{\rm V,0}$    &  $23.1\pm 0.1$    & $23.2\pm 0.1$          & [mag/arcsec$^2$]$^b$    \\
$A_r$   & 0.069     &  0.045         &  \citet{schlafy} (NED)     \\
$A_i$   & 0.052     &  0.033         &   \citet{schlafy} (NED)    \\
$r_{\rm e}$ [arcsec]   &  $4.7\pm 0.2$    & $6.9\pm 0.2$          &  from exponential fit of the i-band profile  \\
$r_{\rm e}$ [kpc]   &  $2.6\pm 0.3$    & $1.3\pm 0.1$          &    \\
PA [deg]           & $40\pm5$ & $25\pm 5$ & eye estimate \\            
b/a & 0.5 & 0.5  & eye estimate \\  
M$_{\rm \star}~[M_{\rm \sun}]$ &  $1.0\times 10^7$ &  $0.9\times 10^7$ & $M/L_i$ from  \citet{rc15}\\       
12+log(O/H)   & $<8.1$     & $8.2\pm 0.2^d$         &  average from N2 and O3N2     \\
              &                   &                      & calibration by \citet{marino} \\                              
12+log(O/H)   &                   & $8.1\pm 0.2^d$         & average from O2 and S2      \\
              &                   &                      & calibration by \citet{pg16} \\
12+log(N/H)   &                   & $6.4\pm 0.2^d$         &                             \\
M$_{\rm HI}$      & $1.2\times 10^9~M_{\rm \sun}$ &              &  \citet{rob04}  \\                                         
\hline
\multicolumn{4}{l}{$^a$ Derived from the H~I radial velocity estimate by \citet{rob04}}\\
\multicolumn{4}{l}{$^b$ Not corrected for extinction.}\\
\multicolumn{4}{l}{$^c$ i- and -r band magnitudes have been transformed into V magnitudes, using the equation ${\rm V=r+0.813(r-i)+0.050}$, valid for ${\rm -0.3\le r-i\le 0.7}$,}\\
\multicolumn{4}{l}{that we derived in the same way as Lupton (2005) from the set of standard stars in the globular cluster NGC~2419 by \citet{stet}.}\\
\multicolumn{4}{l}{$^d$ Weighted mean of the estimates for sources SdI-2~A and SdI-2~B.}\\
\end{tabular} 
\end{center}
\end{table*}

According to the NASA Extragalactic Datasystem NED\footnote{http://ned.ipac.caltech.edu} there are eight galaxies within 1 degree of SdI-1 with recessional velocities in the range 6000~km~s$^{-1}$$\le V_r\le$10000~km~s$^{-1}$. The most remarkable one is the radio galaxy IC~753 that has $V_r=6220$~km~s$^{-1}$ and lies $\simeq 1\degr$ apart in the plane of the sky, corresponding to a projected distance of $\simeq 2.0$~Mpc at the distance of SdI-1.
Two galaxies (SDSS~J115406.35+000158.5 and SDSS~J115254.30-001408.5) have velocities within 200~km~s$^{-1}$ of SdI-1. They are relatively faint (r$\ge 17.7$) and lie at a projected distance of 0.9~Mpc and 1.6~Mpc, respectively. 

There are eighteen galaxies listed in NED within 1 degree of SdI-2 with recessional velocities in the range 1000.0~km~s$^{-1}$$\le V_r\le$4000.0~km~s$^{-1}$, but only three within $\simeq 500$~km~s$^{-1}$ of SdI-2, SDSS~J114428.88-012335.7, SDSS~J114640.78-011749.2, and SDSS~J114325.36-013742.5. All these three are significantly fainter than SdI-2 (r$\ge 20.8$) and lie more than half a degree apart, corresponding to a projected distance larger than 350~kpc at the distance of SdI-2.
In conclusion, both our target dwarfs do not lie in the vicinity of large galaxies and seems to be remarkably isolated.

   \begin{figure}\centering
   \includegraphics[width=\columnwidth]{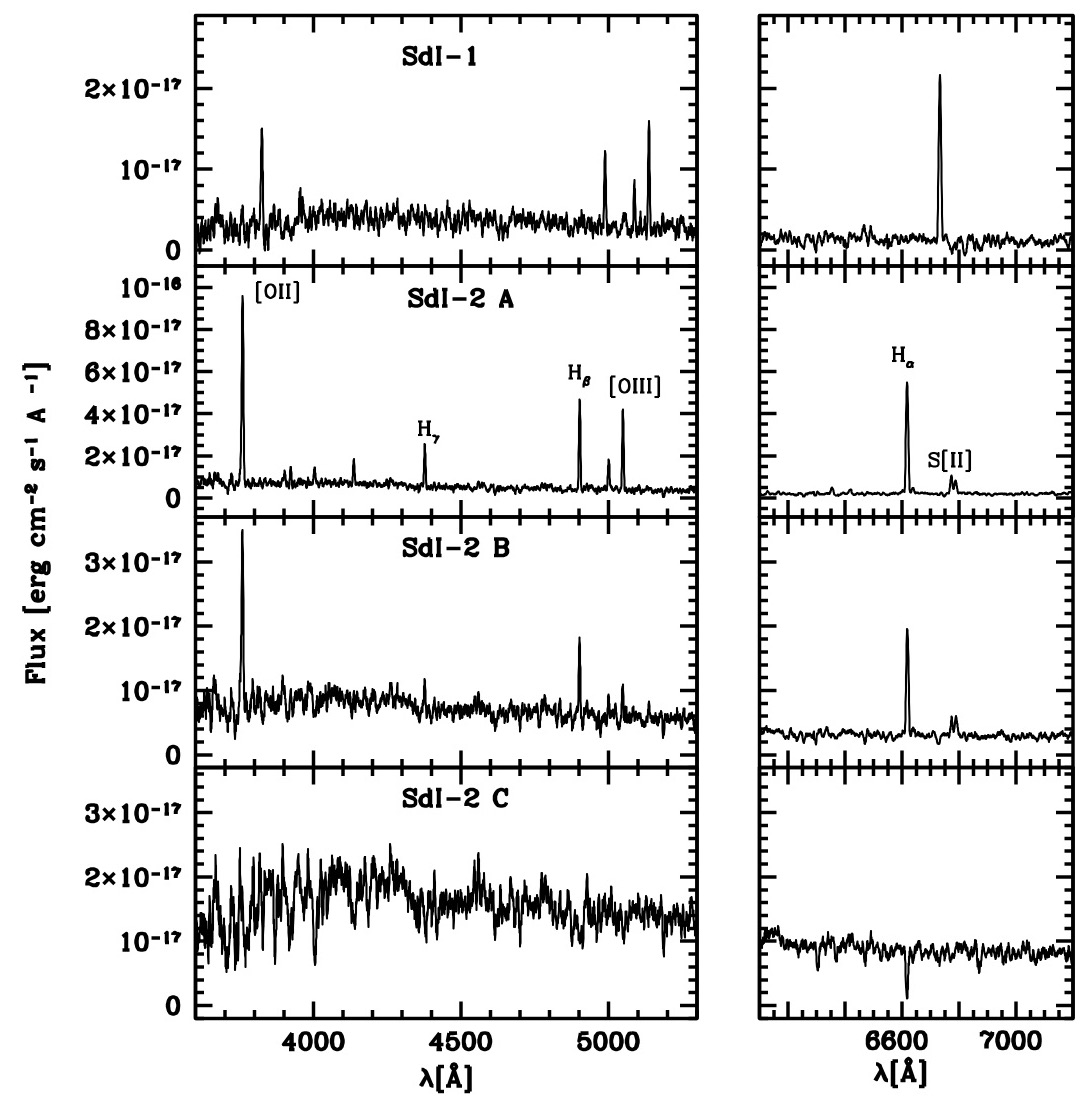}
     \caption{The most significant portions of the spectra from our long-slit observations of SdI1 and SdI2. Upper panel: the single H~II region in SdI-1. Middle panels: the two H~II regions identified in SdI-2. Lower panel: the stellar nucleus of SdI-2. Some noteworthy lines have been labelled in the spectrum of SdI-2~A. }
        \label{allspec}
    \end{figure}


\section{Physical properties of SdI-1 and SdI-2}

In Fig.~\ref{allspec} we show relevant portions of the spectra we obtained from our observations. In SdI-1 this corresponds to a bright and extended H~II region $\simeq 5.2\arcsec$ to the south-west of the center of the galaxy. In SdI-2 we got the spectra of two such H~II regions (sources A and B, $\simeq 5.3\arcsec$ and $\simeq 8.4\arcsec$ to the south-south-west of the center of the galaxy, respectively) and the absorption spectrum of the central nucleus (source C).
To get a rough estimate of the mean age of the nucleus we compared the observed spectrum of source C with a set of low-resolution synthetic spectra of simple stellar populations from the BASTI repository \citep{basti_synt,basti}, finding a satisfactory fit with a model having metallicity Z=0.002 and age=200~Myr.

We estimated the redshift by fitting the centroid of identified atomic lines and averaging the various measures. The uncertainty in the derived radial velocities is dominated by the uncertainty in the zero point which, comparing the velocities from the blue and red spectra, is about 50~km~s$^{-1}$.
The match between our new optical redshift for SdI-1 and the H~I estimate by \citet{rob04} implies that the association between the distant cloud and the stellar counterpart is correct.
Both {\em SdI-1 and SdI-2 have velocities clearly incompatible with association with our GASS candidate mini-haloes}. Since these were the best candidates for counterparts selected from a sample of $\sim 1000$ ICVs, we conclude that it is very unlikely that a local dwarf is associated with these clouds.

We report the radial velocities in the 3K reference frame and then we derive the distance adopting $H_0=73$~km~s$^{-1}$~Mpc$^{-1}$ (the value adopted by NED). We obtain a distance of $112\pm 8$~Mpc and  $40\pm 1$~Mpc for SdI-1 and SdI-2, respectively. 

\subsection{Metallicity}

Using the {\sc iraf} task {\tt splot} we measured the fluxes of recombination lines of H (H$\alpha$ and H$\beta$) and collisional lines of a few ions ([O{\sc ii}], [O{\sc iii}], [N{\sc ii}], [S{\sc ii}], see Table~\ref{linefluxes}). All measured line intensities were corrected for extinction by computing the ratio between the observed and theoretical Balmer decrement for the typical conditions of an H{\sc ii} region 
\citep[see][]{os06}. For the metallicity estimates in SdI-2, described below, we adopt the weighted average of the estimates of the two individual H~II regions, as their abundances are indistinguishable, within the uncertainties.

Due to the absence of electron-temperature diagnostic lines, the gas-phase oxygen abundance of each source was determined with the following {\em strong-line} ratios (also depending on the available lines in each spectrum): 
N2=[N{\sc ii}]/H$\alpha$, O3N2=([O{\sc iii}]/H$\beta$)/([N{\sc ii}]/H$\alpha$), O2=([O{\sc ii}/H$\beta$), and S2=([S{\sc ii}]($\lambda$6717+$\lambda$6730))/H$\alpha$. For N2 and O3N2 we adopt the calibration by \citet{marino}, while for O2 and S2 we adopt the calibration by \citet{pg16}. The final abundance estimates, listed in Table~\ref{mean}, are the average of the values obtained from N2 and O3N2 and the average of the values obtained from O2 and S2. For SdI-2 the two estimates are in good agreement, within the uncertainties.
For SdI-1 we were able to obtain only an upper limit from N2 and O3N2. For both galaxies, the derived abundances are within the range covered by dwarf galaxies of similar luminosity \citep[see, e.g.,][]{lee03}.
The nitrogen-to-oxygen ratio is about log(N/O)$\simeq -1.7$ in SdI-2, indicating that the star formation efficiency is quite high in this galaxy \citep[$\simeq 3-5$~Gyr$^{-1}$,][]{vincenzo}.

\begin{table*}
  \caption{Extinction-corrected fluxes of emission lines (assuming H$_{\rm \beta}$ flux=100.0)}
  \label{linefluxes}
  \begin{tabular}{lccc}
\hline
line               &  SECCO-dI-1 & SECCO-dI-2 A& SECCO-dI-2 B \\
\hline
[OII]$_{\rm 3726+3729}$      & $144.5\pm  9.2$ & $256.2\pm  15.5$ & $252.2\pm 13.4$\\
H$_{\rm \delta}$             & ---             & $25.7\pm  2.1$   & ---            \\
H$_{\rm \gamma}$             & ---             & $46.8\pm  3.1$   & ---            \\
H$_{\rm \beta}$              & $100.0\pm  8.4$ & $100.0\pm 7.8$   & $100.0\pm  5.8$\\
${\rm [OIII]}_{\rm 4959}$    & $46.7 \pm  5.7$ & $27.5\pm  2.2$	  & ---            \\
${\rm [OIII]}_{\rm 5007}$    & $119.5\pm  9.5$ & $40.9\pm  4.9$   & $82.4\pm  4.9$ \\
H$_{\rm \alpha}$             & $289.1\pm 21.7$ & $287.1\pm 14.0$  & $287.0\pm 11.4$\\
${\rm [NII]}_{\rm 6584}$     & $<6.0 $         & $10.6\pm  3.2$   & $15.5\pm  1.4$ \\
${\rm [SII]}_{\rm 6717}$     & ---             & $18.1\pm  3.5$   & $41.8\pm  2.3$ \\
${\rm [SII]}_{\rm 6731}$     & ---             & $38.3\pm 4.3$    & $31.3\pm  2.0$ \\
\hline
 [$10^{17}$ erg~cm$^{-2}$~s$^{-1}$] & $5.9\pm0.5$           &    $24.5\pm1.4$       & $7.1\pm0.5$            \\
\hline
\end{tabular} 
\end{table*}

\subsection{Surface Photometry}

\begin{figure}\centering
   \includegraphics[width=\columnwidth]{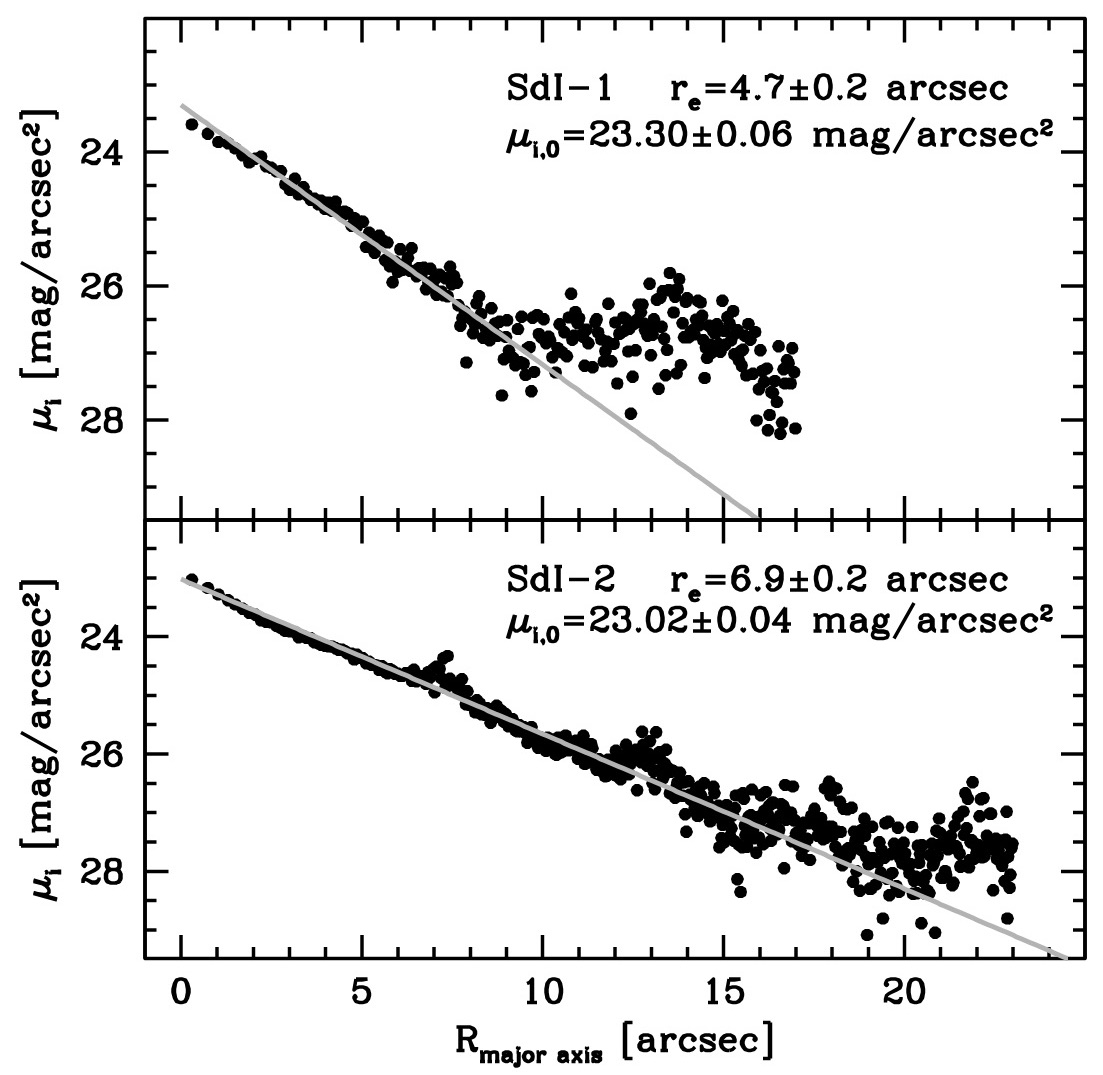}
     \caption{i-band surface brightness profiles of SdI-1 (upper panel) and SdI-2 (lower panel) from surface photometry on elliptical apertures (with b/a and PA as reported in Tab.~\ref{mean}. The grey lines are exponential profiles that best fit the observed profiles. The best-fit estimates of $r_{\rm e}$ and $\mu_{\rm i,0}$ are also reported.}
        \label{profi}
    \end{figure}

In Fig.~\ref{profi} we show the i-band surface brightness profiles of our target galaxies, obtained by photometry on elliptical apertures performed with the APT software \citep{apt}. The axis ratio and position angle were estimated by eye, superposing ellipses to the images of the galaxies. r-band profiles were obtained in the same way and are very similar in shape, but we prefer to analyse i-band profiles because they should be minimally affected by the light of bright H~II regions. Indeed slightly larger effective radii are obtained from r-band profiles.
 
In spite of some noise due to the presence of fore/background sources or compact sources within the galaxies, the overall profiles emerge very clearly, and in both cases they can be satisfactorily fitted by a simple exponential model. 
Contamination from nearby but unrelated sources especially affects the outer part of the profile of SdI-1, which flattens beyond $R=10\arcsec$ mainly due to the contribution from the possible companion lying to the west of the galaxy (see Sect.~\ref{targ}). We limited our fit to the nearly uncontaminated inner regions to avoid overestimating the size of the galaxies.
The effective radii ($r_{\rm e}$) along the major axis, the central surface brightness ($\mu_{\rm V,0}$) and the associated uncertainties have been estimated by performing a linear regression on the points within $R=10.0\arcsec$, for SdI-1, and within $R=16.0\arcsec$, for SdI-1, with the macro {\tt lm} of the package {\sc R}\footnote{ \tt www.r-project.org}. This simple approach seems fully adequate in this context, is straightforward and provides robust estimates of these parameters.

We note that our $r_{\rm e}$ value for SdI-1 is in reasonable agreement with the estimate by \citet{rob04} from much shallower images. To minimise the effect of fore/background sources we derived the integrated magnitudes and the mean surface brightness within $r_{\rm e}$ 
from the best fitting exponential profiles, using the equations provided by \citet{GD05}. All the photometric and structural parameters derived in this way are listed in Table~\ref{mean}.

The ${\rm (r-i)_0}$ colors of SdI-1 and SdI-2 ($-0.1$ and $+0.1$, respectively) are typical of star-forming galaxies and on the blue side of the range spanned by blue UDGs from \citet[][see Sect.~4 for a comparison with these galaxies]{rt16}. 
Following these authors we used the relations by \citet{rc15} to derive a rough estimate of the stellar mass from the integrated i-band luminosity and the $(r-i)_0$ colors; both galaxies have $M_{\rm \star}\simeq 1\times10^7~M_{\rm \sun}$ (see Tab.~\ref{mean}). For SdI-1 we have also a reliable estimate of the H~I mass from \citet{rob04} that allows us to compute the ratio of the H~I mass to V-band luminosity (from Tab.~1) to be $\frac{M_{\rm HI}}{L_V}\simeq 6.2$, larger than the typical values of gas-rich dwarfs in the Local Group \citep[$\frac{M_{\rm HI}}{L_V}\sim 2$;][]{mcc}. The ratio of H~I mass to stellar mass, $\frac{M_{\rm HI}}{M_{\rm \star}}\simeq 90$ places SdI-1 in the realm of {\em almost-dark} galaxies \citep[see][]{almostdark,secco_muse}.

\section{A link between dwarf irregulars and UDGs?}
\label{conc}

Having at our disposal newly derived structural parameters and distances of our target galaxies, we noted that they have absolute magnitude, surface brightness (see Table~\ref{mean}), and, in particular,
physical sizes ($r_{\rm e}=2.6$~kpc and $r_{\rm e}=1.3$~kpc, for SdI-1 and SdI-2, respectively) in the range covered by the recently identified class of ultra diffuse galaxies\footnote{There is not yet a generally accepted definition of the class, hence similarity must be intended in a broad sense. For example, SdI-1 fulfills the main size criterion ($r_{\rm e}>1.5$~kpc) adopted by \citet{yagi} in their definition of UDGs, independently of adopting the {\em major axis} or the {\em circularised} effective radius, SdI-2 fails by a small amount, and both galaxies fail to fulfill the $\mu_{\rm V,0}>24.0$~mag/arcsec$^2$ criterion. However both galaxies are consistent with {\em all} the \citet{yagi} criteria once the fading by passive evolution is taken into account (see below). Moreover, both SdI-1 and SdI-2 overlap with Fornax UDGs identified by \citet{munoz} in luminosity, radius and $\langle \mu\rangle_{\rm e}$.} \citep[UDG,][]{udg}.

UDGs are roundish amorphous galaxies having ``...the sizes of giants but the luminosity of dwarfs...'' \citep{beasley} that have been recently discovered in large numbers in clusters of galaxies \citep[see][and references therein]{udg,koda,mihos,munoz,burg}, with some examples in other environments \citep[see][and references and discussion therein]{david,yagi}. In general UDGs lie on the red sequence of galaxies and their light profiles are well approximated by exponential laws \citep{koda,yagi}. The fact that they survive in dense environments without obvious signs of tidal distortions, as well as the first analyses of the kinematics of individual UDGs, strongly suggest that they are dark matter dominated systems \citep{udg,burg,udg3,beasley,zar}. On the other hand it is still not established if, e.g., they are failed giant galaxies \citep{udg2} or extended and quenched dwarfs \citep{beas2}.

   \begin{figure*}\centering
   \includegraphics[width=\textwidth]{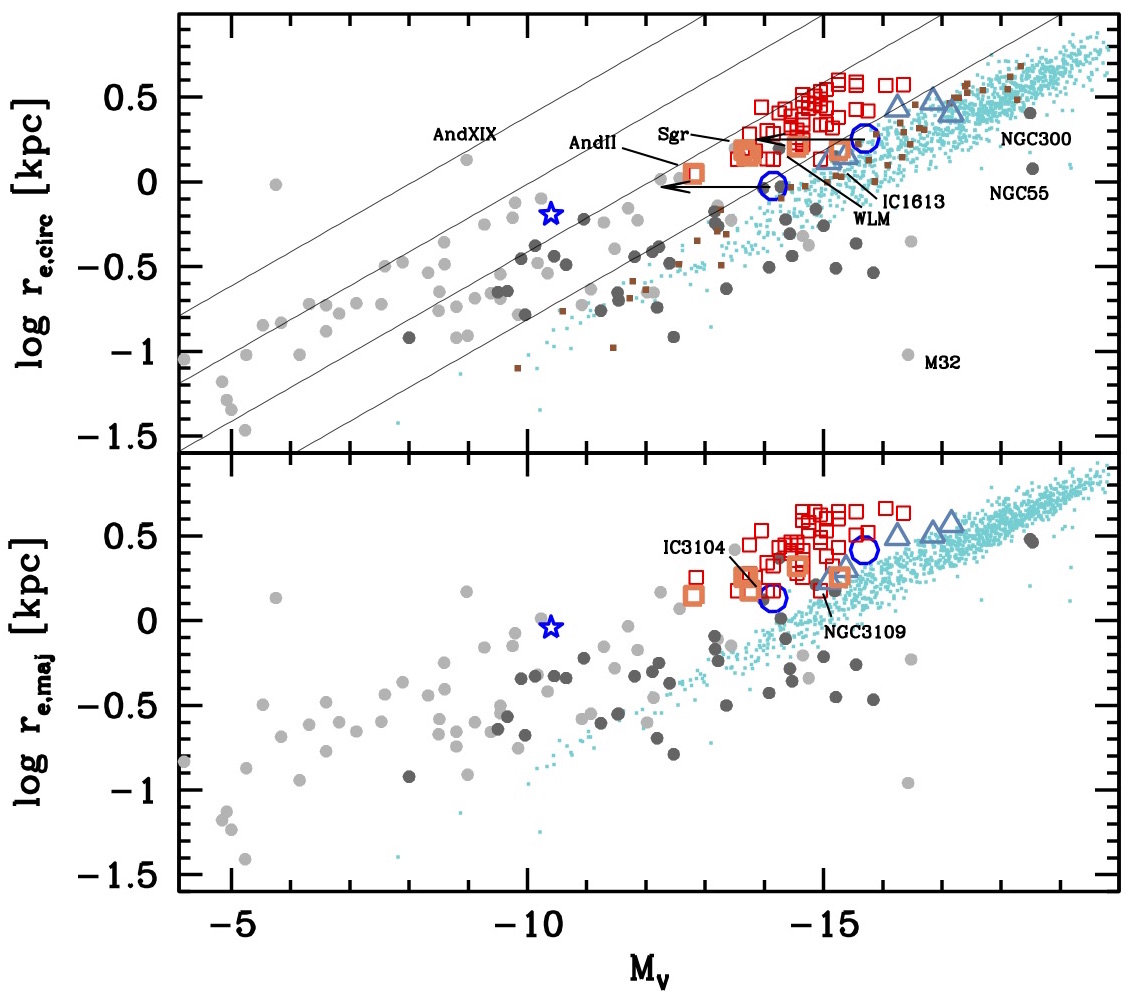}
     \caption{Comparison between SdI-1 and SdI-2 (open blue circles), local dwarf galaxies from the most recent version of the catalog by \citet[][grey filled circles; darker points are gas-rich galaxies, having $M_{\rm HI}/L_V>0.1$, in solar units]{mcc}, and UDGs from \citet[][red empty squares]{udg} in the $M_V$ vs. effective radius plane. 
The arrows show the fading of a metal-poor single-burst stellar system (with the size and luminosity of SdI-1 and SdI-2) passively evolving from age=0.5~Gyr to age=6.0~Gyr, according to BASTI evolutionary models \citep{basti_synt}. 
Thick orange open squares are red UDGs and thick azure open triangles are blue UDGs from \citet{rt16}.
Turquoise points are gas-rich low SB galaxies from \citet{du}. Small brown filled squares are blue diffuse dwarfs (BDDs) from \citet[][not plotted in the lower panel since only $r_{\rm e,circ}$ values were available]{bdd2}. The blue open star is the almost-dark star forming galaxy SECCO~1, from \citet{secco1} and \citet{secco_muse}. The point labelled as Sgr  is the Sagittarius dwarf spheroidal galaxy. 
In the lower panel the effective radius measured along the major axis is adopted ($r_{\rm e,maj}$) while in the upper panel the circularised effective radius ($r_{\rm e,circ}=r_{\rm e,maj}\sqrt{b/a}$) is used. Some noteworthy galaxies have been labelled, for reference. In the upper panel, four $\langle \mu_V\rangle_{\rm e} ={\rm const.}$ lines are also plotted, for reference, at $\langle \mu_V\rangle_{\rm e} =30.5, 28.5, 26.5, 24.5$~mag/arcsec$^2$, from top to bottom. }
        \label{UDG}
    \end{figure*}


In Fig.~\ref{UDG} we compare the sample of Coma cluster UDGs by \citet{udg} with SdI-1, SdI-2 and the dwarf galaxies in the Local Volume, from the compilation by \citet{mcc}, in the absolute magnitude vs. effective radius plane. 
Local dIrr galaxies are plotted in a darker tone of grey with respect to dSphs and dwarf ellipticals. NGC~300 and NGC~55 have been plotted for reference although they do not fit the definition of dwarf galaxies. To expand our view we included also the BDDs of \citet{bdd,bdd2} and the H~I-selected sample of gas-rich LSB galaxies within $\simeq 250$~Mpc by \citet{du}. These authors derived new, more reliable estimates of integrated magnitudes and effective radii by re-analysing SDSS images. Absolute g- and 
r-band integrated magnitudes from \citet{du} were transformed into V-band magnitudes by using Lupton (2005)\footnote{http://www.sdss3.org/dr9/algorithms/sdssUBVRITransform.php} equations.
Also g-band magnitudes of \citet{udg} galaxies were converted into V-band  with the same equation,  adopting the mean color of red sequence galaxies at that luminosity, ${\rm (g-r)_0=0.6}$, from \citet{blanton}. UDGs from \citet[][see discussion below]{rt16} have been included in Fig.~\ref{UDG} with the same transformation, adopting mean colors of ${\rm (g-r)_0=0.25}$ and ${\rm (g-r)_0=0.6}$, for blue and red UDGs, respectively (see their Fig.~4).
In Fig.~\ref{UDG} we show two slightly different versions of the plot. In the lower panel we use the effective radius measured along the major axis of the galaxies ($r_{\rm e,maj}$), while in the upper panel we use the circularised radius ($r_{\rm e,circ}=r_{\rm e,maj}\sqrt{b/a}$), since this has been adopted in several comparisons of the same kind including UDGs \citep[see, e.g.,][]{beasley,udg2}.

We note that, {\em independently of the version of the plot, not only SdI-1 and SdI-2, but also a few local large dwarf irregulars partly overlap the distribution of UDGs} (WLM, IC~1613, in particular, but also NGC~3109 and IC~3104, if $r_{\rm e,maj}$ is considered). Indeed, in their extensive literature search for previously identified UDGs, \citet{yagi} found six galaxies satisfying {\em all} their UDG criteria in the catalog of nearby dwarf irregulars by \citet{HE06}, concluding that an overlap between the two classes may indeed exist.

The adoption of $r_{\rm e,circ}$ makes the overlapping range slightly narrower, due to the fact that UDGs have, on average, a much rounder shape than dIrrs.
It is also important to stress that while it was generally recognised that in the $M_V$ vs. log~$r_{\rm e}$ plane, UDGs lie at the tip of the sequence of dSphs and dIrrs, the inclusion of the latter class of galaxies in the plot (usually not performed) makes the overall distribution much more continuous: there seems to be no gap between dSphs+dIrrs and UDGs. Expanding the view to non-local systems suggests that large dIrrs have indeed effective radii in the range 1-4~kpc \citep[see, e.g.,][their Fig.~9, in particular]{lange}. 

The sharp edge of the distribution of \citet{du} galaxies toward low SB values in Fig.~\ref{UDG} strongly suggests that the lack of a more substantial overlap with the UDGs may be merely due to incompleteness: a population of star-forming dwarfs with $\langle \mu_V\rangle_{\rm e}>24.5$~mag/arcsec$^2$ may still be waiting to be uncovered by future more sensitive surveys\footnote{It is interesting to note that \citet{yagi} stated that the data-reduction strategy they adopted implies a bias against the detection of extended dwarf irregulars.}. This hypothesis is confirmed by the fact that the sample of BDDs by \citet{bdd2}, which has been selected from the same source (SDSS images), shows the same cut in surface brightness. Fig.~\ref{UDG} shows also that \citet{du} and \citet{bdd2} galaxies and local dIrrs are indistinguishable in this plane.

We do not know if the similarity between large dIrrs and UDGs\footnote{The similarities include also the typical shape of the light profiles (nearly exponential), the incidence of stellar nuclei \citep[][]{yagi}, and the  presence of  globular cluster systems \citep{beasley,beas2,udg3}. Note that at least one UDG has been found hosting a globular cluster population significantly larger than typical dwarfs of the same luminosity \citep{udg3}.} is hinting at an evolutionary link between the two classes, but it is certainly worth noting, since a clear explanation for the origin of UDGs is still lacking and they can also be the end-product of different evolutionary channels \citep{zar}. For instance, it seems to support the hypothesis that UDGs may be ``{\em quenched Large Magellanic Cloud-like systems}'', recently put forward by \citet{beas2}. It may be conceived that the {\em tidal stirring} process that is supposed to transform small gas-rich disc dwarfs into dSph around Milky Way-sized galaxies \citep{lucionat}, acting on a larger scale, can also transform large dIrrs into UDGs within galaxy clusters, removing the gas, stopping the star formation and redistributing the stars into an amorphous spheroid (see also \citealt{amorisco} for a possible relation between disc galaxies and UDGs, and \citealt{burk} 
for shape arguments not supporting this relation). Interestingly, the only Local Group
quiescent galaxy overlapping the distribution of UDGs in Fig.~\ref{UDG} \citep[see also][]{yagi}, the Sagittarius dSph, is believed to have evolved to its present amorphous and gas-less state from a star-forming galaxy of mass similar to the Small Magellanic Cloud, mainly driven by the tidal interaction with the Milky Way that is disrupting it \citep{maj,NO}. 
However, it must be noted, in this context, that typical UDGs do not shows signs of ongoing disruption.
In any case, the continuity of the sequence including dIrrs of any size and dwarf spheroids of any size (dSphs and UDGs) in the $M_V$ vs. log~$r_{\rm e}$ plane may suggest similar progenitors for all these LSB systems over more than six orders of magnitude in luminosity. 

In this context, the very recent work by \citet[][RT16 hereafter]{rt16} is particularly relevant. These authors identified a population of possible progenitors of UDGs ({\em blue} UDGs, from their colors significantly bluer than classical {\em red} UDGs) in the outskirts of galaxy groups containing classical UDGs. RT16 demonstrated that a few Gyrs of passive evolution would
transform their blue UDGs into classical red ones, by reddening their colors, fading their surface brightness and total luminosity while keeping their large size nearly unchanged. SdI-1 and SdI-2 have size, stellar mass, luminosity, surface brightness, ellipticity and color very similar to the RT16 blue UDGs (see, e.g., Fig.~\ref{UDG}), hence RT16 results on the evolutionary path of blue UDGs applies also to our galaxies as well as to other local dIrrs plotted in Fig.~\ref{UDG}. In particular, according to canonical solar-scaled BASTI\footnote{\tt http://basti.oa-teramo.inaf.it} stellar evolutionary models \citep{basti_synt,basti} for a simple stellar population with metallicity Z=0.002, the passive evolution from an age=0.5~Gyr to age=6.0~Gyr would led to a fading by 1.84 magnitudes in V-band, driving the central surface brightness of SdI-1 and SdI-2 down to $\mu_{\rm V,0}\simeq 25.0$~mag/arcsec$^2$, fully in the realm of classical UDGs, as displayed by the arrows plotted in the upper panel of Fig.~\ref{UDG}. 

SdI-1 and SdI-2 seem to be even more isolated than their RT16 siblings, since they lie at more than 900~kpc and 350~kpc from their nearest known neighbours, respectively, while RT16 blue UDGs are within 250-550~kpc from the centre of the galaxy groups they are associated to. Moreover their nearest neighbours are dwarf galaxies, while the groups where RT16 UDGs live host also giant galaxies. {\em SdI-1 and SdI-2  are perhaps the most isolated UDGs (or UDG progenitors) identified until now, indicating that isolated field dwarfs can indeed evolve into UDGs.} 
Hence, SdI-1 and SdI-2 lend additional and independent support to the scenario proposed by \citet{rt16}, in which the progenitors of classical UDGs were dwarfs born in the field and then processed within galaxy groups and, finally, in galaxy clusters. 

The properties of SdI-1 and SdI-2 also fit nicely with those of the dwarfs identified by \citet{dicint} as counterparts of UDGs in cosmological simulations including feedback processes. Also in the \citet{dicint} scenario UDGs are born as dwarf galaxies, but their evolution is driven by internal processes (feedback-driven gas flows, in particular), hence it is independent of the environment and it is expected to take place also in the field.  
A specific prediction of the simulations by \citet{dicint} is that UDGs evolved in isolation should have larger gas content than regular dwarfs of similar stellar mass. It is intriguing to note that this prediction is vindicated by the the very high H~I/$M_{*}$ ratio observed in SdI-1, the most extended among our two isolated, star-forming UDGs.

\section*{Acknowledgements}
We are grateful to an anonymous referee for a very careful reading of the original manuscript and for useful suggestions that allowed us to make a more thorough analysis. 
This research has made use of the NASA/IPAC Extragalactic Database (NED) which is operated by the Jet Propulsion Laboratory, California Institute of Technology, under contract with NASA. This research has made use of the SIMBAD database, operated at CDS, Strasbourg, France.







\bsp	
\label{lastpage}
\end{document}